%% file: Mixture.tex
\documentclass[%
aps,
pra,
reprint,
twocolumn,
superscriptaddress,
longbibliography
]{revtex4-2}
\usepackage{amsmath}% need to load this extra since revtex loads trash
\usepackage{mathtools} % for coloneqq
\usepackage{graphicx}% Include figure files
\usepackage{dcolumn}% Align table columns on decimal point
\usepackage{bm}% bold math
\usepackage{hyperref}% add hypertext capabilities
\usepackage{booktabs}% add hypertext capabilities
\hypersetup{% define custom hyper reference setup
	colorlinks=true,
	urlcolor=blue,
	citecolor=blue,
	linkcolor=blue, % set the link color
	linktocpage=true, % makes the page number as hyperlink in table of content
	linktoc=all,     % set to all if you want both sections and subsections linked
}
\usepackage[mathlines]{lineno}
\usepackage{siunitx} % for units
\usepackage{comment} % to comment out larger portions of the .tex document
\usepackage{braket} % for bras and kets

\begin{document}
\title{Reconfigurable Dysprosium Isotope Mixtures with Selectable Quantum Statistics}

\author{Zibin Jiang}%
\thanks{These authors contributed equally to this work.}
\affiliation{Department of Physics and Guangdong Basic Research Center of Excellence for Quantum Science, Southern University of Science and Technology, Shenzhen 518055, China}%

\author{Kaiyue Wang}%
\thanks{These authors contributed equally to this work.}
\affiliation{Department of Physics and Guangdong Basic Research Center of Excellence for Quantum Science, Southern University of Science and Technology, Shenzhen 518055, China}%

\author{Shenshuang Nie}
\affiliation{Department of Physics and Guangdong Basic Research Center of Excellence for Quantum Science, Southern University of Science and Technology, Shenzhen 518055, China}%
\author{Junrong Huang}%
\affiliation{Department of Physics and Guangdong Basic Research Center of Excellence for Quantum Science, Southern University of Science and Technology, Shenzhen 518055, China}%

\author{Fucheng Qin}%
\affiliation{Department of Physics and Guangdong Basic Research Center of Excellence for Quantum Science, Southern University of Science and Technology, Shenzhen 518055, China}%

\author{Mingyang Guo}%
\email{guomy@sustech.edu.cn}
\affiliation{Department of Physics and Guangdong Basic Research Center of Excellence for Quantum Science, Southern University of Science and Technology, Shenzhen 518055, China}%
\affiliation{Quantum Science Center of Guangdong-Hong Kong-Macao Greater Bay Area, Shenzhen 518045, China}

\date{\today}

\begin{abstract}

Dipolar quantum mixtures offer new opportunities to explore many-body phenomena shaped by long-range anisotropic interactions, intercomponent coupling, and quantum statistics. Here we develop a reconfigurable platform for bosonic and fermionic dysprosium isotope mixtures using a single-isotope-like optical layout. We time-multiplex the 421-nm light to generate isotope-selective slow-atom fluxes, while spectrally engineered 626-nm light provides simultaneous common-path narrow-line cooling and trapping. By reconfiguring optical frequencies, we prepare seven isotope pairs spanning Bose--Bose, Bose--Fermi, and Fermi--Fermi combinations, with continuous population-balance control through the 421-nm dwell-time allocation. Systematic comparisons with sequence-matched single-isotope references, from magneto-optical-trap loading and compression to optical-dipole-trap transfer, reveal nearly single-isotope-like performance for most combinations. The larger deviations occur only for certain isotope pairs and preparation stages, with controlled measurements revealing optical sensitivities during loading and additional losses under high-density conditions. Our results make isotope composition and quantum statistics selectable within a strongly dipolar system while largely preserving single-isotope preparation simplicity and performance, opening new possibilities for exploring multicomponent dipolar quantum matter.

\end{abstract}

\maketitle

\section{Introduction}
Magnetic atoms have established long-range and anisotropic interactions as an important ingredient in ultracold quantum matter~\cite{Lahaye2009,Norcia2021b,Bottcher2021,Chomaz2023}. The dipole--dipole interaction has enabled quantum droplets stabilized by beyond-mean-field fluctuations~\cite{Kadau2016,Schmitt2016,Ferrier-Barbut2016,Chomaz2016}, roton softening as a precursor to density instabilities~\cite{Santos2003,Petter2019,Hertkorn2021}, and supersolid phases with crystalline order coexisting with superfluid stiffness~\cite{Tanzi2019,Bottcher2019,Chomaz2019,Guo2019,Natale2019a,Sohmen2021}. Further control has been achieved by imposing reduced dimensions~\cite{Kao2021,Yang2024,He2025,He2025a}, optical lattices~\cite{Baier2016,Su2023a} and bilayer geometries~\cite{Du2024}, which introduce new confinement conditions and characteristic length scales. 

Composition provides a qualitatively different control axis by introducing a relative-density degree of freedom in addition to the total density. These coupled channels give rise to interaction-driven miscibility and phase separation~\cite{Hall1998,Papp2008a,Tojo2010,Wang2015}, coupled collective modes, impurity physics~\cite{Schirotzek2009,Jorgensen2016}, and mediated interactions~\cite{DeSalvo2019,Edri2020}. In dipolar mixtures, long-range and anisotropic interactions introduce additional nonlocal and direction-dependent coupling between density and composition responses, modifying stability and collective excitations, and also supporting composite droplets and coupled spatial order~\cite{Saito2009,Wilson2012,Bland2022,Li2022,Halder2023,Scheiermann2023}. This interplay is particularly rich in Bose--Fermi mixtures, where the fermionic density response and mediated interactions can acquire pronounced momentum and angular dependence, leading to phenomena with no direct analogue in contact-interacting mixtures~\cite{Kain2013,Adhikari2013,Lewkowicz2025}.

Accessing this broader landscape requires strongly dipolar mixtures with selectable quantum statistics, tunable composition, and strong spatial overlap. Experimentally, multicomponent dipolar gases can be realized through several complementary approaches. Heteronuclear lanthanide mixtures, exemplified by erbium--dysprosium, combine two strongly magnetic species and have established an important route to binary dipolar gases, while generally requiring species-dependent cooling and trapping conditions~\cite{Ilzhofer2017,Trautmann2018}. Spin or internal-state mixtures of a single dipolar species provide a complementary limit, with identical mass and trapping conditions, and recent progress has shown that dipolar relaxation can be suppressed in suitable regimes~\cite{Baier2018a,Barral2024,Lecomte2025}. Isotope mixtures offer a distinct route~\cite{Lu2012,Nie2026,Durbeck2026,Durbeck2026a}, combining closely matched single-particle properties with optical distinguishability through isotope shifts, thereby enabling component-selective manipulation with low crosstalk.

Dysprosium is particularly well suited to this isotopic approach. Its isotopes share nearly identical electronic structure, magnetic moment and optical polarizability, and therefore experience nearly common trapping potentials and gravitational sag. At the same time, well-resolved isotope shifts on the rich broad- and narrow-line optical transitions provide spectral selectivity for isotope-selective cooling, manipulation and detection. The availability of several stable bosonic and fermionic isotopes further makes it possible to realize Bose--Bose, Bose--Fermi, and Fermi--Fermi mixtures within the same atomic platform. Together, these advantages make dysprosium isotope mixtures a particularly versatile setting for multicomponent dipolar quantum gases. 

Here we develop a reconfigurable preparation scheme for bosonic and fermionic dysprosium isotope mixtures while retaining a single-isotope-like optical layout. Our approach combines time-multiplexed 421-nm light for isotope-selective Zeeman slowing and transverse cooling with spectrally engineered 626-nm light for simultaneous narrow-line magneto-optic trap (MOT) operation. By reconfiguring the optical frequencies, we prepare seven isotope pairs covering Bose--Bose, Bose--Fermi, and Fermi--Fermi mixtures, together with continuous population-balance control through the 421-nm dwell-time allocation. To evaluate the performance of this common preparation pathway, we introduce a hierarchy of sequence-matched single-isotope controls that separately assess the effects of dual-frequency 626-nm operation, interrupted 421-nm cooling, and full dual-isotope preparation. These comparisons extend from MOT loading and compression to optical-dipole-trap (ODT) transfer. We find that most components retain nearly single-isotope-like preparation performance, while stage-resolved measurements and targeted controls reveal pair-dependent optical sensitivities and additional high-density losses in certain isotope combinations. Together, these results establish a systematically validated and reconfigurable preparation pathway for extending dysprosium isotope mixtures toward quantum-degenerate dipolar mixtures with controllable isotope composition, population balance, and quantum statistics.

\section{Experimental scheme and optical isotope addressing}

\subsection{Experimental setup and scheme}

\begin{figure*}[t!]
	\includegraphics[width=1.8\columnwidth]{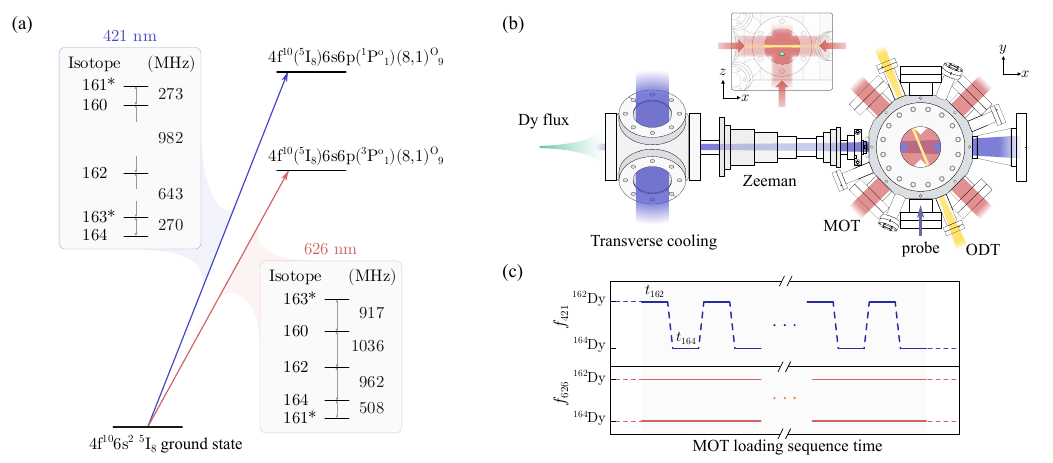}
	\caption{\textbf{Experimental scheme and reconfigurable isotope-mixture platform.}
		(a) Relative isotope shifts of the 421-nm broad-line and 626-nm narrow-line transitions for the Dy isotopes used in this work. The \(\Delta J=+1\) cycling transitions from the \(J=8\) ground fine-structure level are shown. For the fermionic isotopes with additional hyperfine structure, indicated by the asterisks, the \(F=21/2\rightarrow23/2\) transition is used. 
		(b) Top view of the experimental setup. The inset shows a side view of the science chamber along the side-imaging direction ($y$), with the five-beam 626-nm MOT located well below the geometric center of the chamber.
		(c) Experimental sequence for dual-isotope operation, shown for the $^{162}$Dy--$^{164}$Dy combination. During MOT loading, the 421-nm light alternates between the selected isotopes with programmable dwell times $t_i$ for isotope $i$, while the 626-nm light simultaneously addresses both components. The dual-isotope 626-nm configuration is maintained through the CMOT stage before ODT transfer.
	}
	\label{fig:setup}
\end{figure*}

The central design principle of our experiment is to preserve the simplicity of a single-isotope dysprosium apparatus while making isotope selectivity programmable through optical-frequency control. The closely matched single-particle properties of Dy isotopes allow them to share a common cooling and trapping geometry, while the well-resolved isotope shifts of the 421-nm and 626-nm transitions enable isotope-resolved optical addressing [Fig.~\ref{fig:setup}(a)]. This design allows different isotope pairs to be prepared with single-isotope-like beam geometry and preparation procedures, from MOT loading to ODT transfer.

Our isotope-mixture preparation builds upon a single-isotope experimental setup, as shown in Fig.~\ref{fig:setup}(b). Dysprosium atoms emitted from a high-temperature effusive oven form an atomic beam that is transversely cooled and longitudinally slowed on the broad 421-nm transition. The slowed atoms are captured in the science chamber by a five-beam narrow-line MOT operating on the 626-nm transition~\cite{Ilzhofer2017}, whose natural linewidth is about 136~kHz. After MOT loading, the atoms are further cooled and compressed in a compressed-MOT (CMOT) stage and subsequently transferred into an acousto-optic-deflector-scanned single-beam ODT.

The distinct roles of the two cooling transitions motivate us to adopt complementary isotope-addressing schemes, as illustrated in Fig.~\ref{fig:setup}(c). Before MOT capture, the 421-nm light provides transverse cooling and Zeeman slowing of the atomic beam. Since this stage prepares the incoming slow atomic flux rather than maintaining a trapped sample, the selected isotope pair can be addressed sequentially by switching the 421-nm frequency between the corresponding isotope resonances. By contrast, the 626-nm light provides the damping and restoring forces of the narrow-line MOT. To maintain uninterrupted confinement of both isotopes, we generate two frequency components that simultaneously address their respective 626-nm MOT transitions within a common optical path. The dual-isotope 626-nm configuration is maintained through the subsequent CMOT stage, during which the 421-nm beams are turned off while the 626-nm MOT spectrum and trapping parameters are ramped for further cooling and compression.

Beyond isotope-pair selection, the time-multiplexed 421-nm sequence provides an additional control over the mixture composition through the relative dwell-time allocation during MOT loading. In each 1-s switching cycle, the 421-nm light dwells on isotope $i$ for a time $t_i$. After $n$ switching cycles, the accumulated effective loading time for isotope $i$ is $T_i^{421}=n t_i$, which is used below to characterize the isotope-resolved loading dynamics. For a selected isotope pair $\{i,j\}$, we define the 421-nm dwell-time bias as
$\beta_{i,j}=({t_{j}-t_{i}})/({t_{i}+t_{j}})$,
such that positive and negative values favor the incoming fluxes of isotope $j$ and $i$, respectively. Varying $\beta_{i,j}$ tunes the relative loading of the two isotopes while leaving the common 626-nm MOT and CMOT parameters unchanged. The isotope pair is therefore selected through the addressed optical frequencies, whereas the relative population is controlled through the loading bias $\beta_{i,j}$. The implementation and characterization of these complementary addressing schemes are described in the following subsections.

\begin{figure*}[tb!]
	\includegraphics[width=1.8\columnwidth]{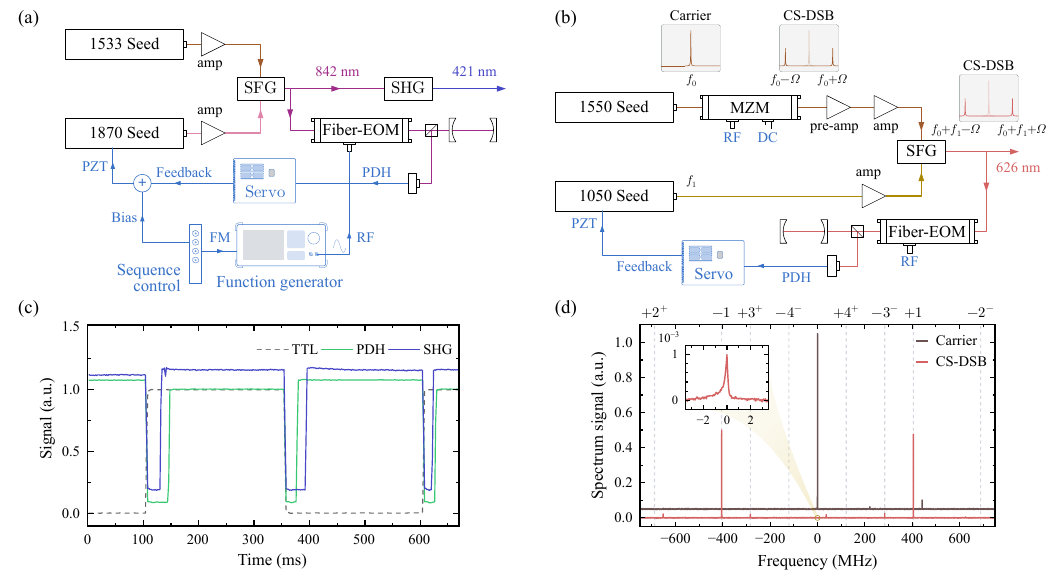}
	\caption{\textbf{Implementation and characterization of isotope-selective 421-nm and 626-nm addressing.}
	(a) \& (c) Time-multiplexed 421-nm addressing. 
	(a) Schematic of the frequency switching, in which the EOM RF drive is switched between targeted isotopes together with synchronized automatic bias compensation for laser frequency and doubling cavity locks.
	(c) Representative switching traces of the PDH error signal (green) and doubling-cavity transmission (blue) synchronized with the TTL signal (gray), showing reproducible relocking within 70~ms.
	(b) \& (d) Spectrally engineered 626-nm addressing. 
	(b) Spectral engineering of the 626-nm MOT laser using a fiber MZM operated in the CS-DSB regime generates the isotope-resolved frequency components on the 1550-nm seed before sum-frequency generation. 
	(d) ULE cavity transmission spectra of the CS-DSB 626-nm light with 405~MHz RF modulation (red) and the unmodulated carrier (brown, vertically offset). The residual carrier is suppressed by over 30~dB (inset), with more than 96\% of the optical power concentrated in the targeted first-order sidebands. The spectrum is folded into a single cavity free spectral range (FSR), with dashed lines marking the expected positions of the labeled sideband orders. Superscripts $\pm$ indicate folding from the neighboring FSRs indexed by $\pm1$.
	}
	\label{fig:imple}
\end{figure*}

\subsection{Time-multiplexed 421-nm cooling}

In our apparatus, 421-nm light is produced by sum-frequency generation of 1870-nm and 1533-nm fiber lasers to 842~nm, followed by second-harmonic generation in a resonant doubling cavity~\footnote{The 421-nm laser system is based on a commercial frequency-conversion platform supplied by Preci Laser.}. The cavity-enhanced conversion provides the optical power required for transverse cooling and Zeeman slowing, but the resonant doubling cavity favors operation at a single optical frequency. Frequency components separated by the isotope shifts therefore cannot be simultaneously enhanced with comparable efficiency. We instead implement isotope selectivity on the 421-nm transition in the time domain, sequentially addressing the selected isotope resonances without duplicating the broad-line cooling optics.

The frequency switching is implemented before frequency doubling, as shown in Fig.~\ref{fig:imple}(a). The 842-nm light is stabilized to an ultralow-expansion (ULE) reference cavity with a resonance linewidth of approximately 0.5~MHz. A broadband fiber electro-optic modulator (EOM) provides a programmable radio-frequency (RF) offset relative to the cavity reference. Switching this RF drive changes the cavity-referenced 842-nm frequency and, after frequency doubling, alternates the 421-nm light between the selected isotope resonances within the same transverse-cooling and Zeeman-slower beam paths.

During MOT loading, the 421-nm frequency repeatedly alternates between the two isotope settings according to the programmed dwell times. The light is actively extinguished during each frequency jump and relocking interval to avoid scattering from transient intermediate frequencies. Reliable time multiplexing therefore requires reproducible recovery of both the laser-frequency lock and the resonant doubling-cavity lock after each jump. We achieve this using an automatic bias-compensation protocol that synchronously adjusts the control parameters of the frequency-locking and doubling-cavity feedback loops. With optimized servo parameters, both locks recover within 70~ms~ [Fig.~\ref{fig:imple}(c)], mainly limited by the present bias-ramp time of about 50~ms. This recovery time is substantially shorter than the typical 500-ms dwell interval, keeping the switching dead time small and enabling stable routine operation over more than one month.

The same frequency agility is also used for component-resolved detection on the broad 421-nm transition. For imaging, the probe light is tuned to one isotope resonance at a time, while the other isotope is detuned by the isotope shift, far exceeding the transition linewidth. This provides straightforward low-crosstalk selection of the imaged component. Thus, the time-multiplexed 421-nm system supports isotope-selective broad-line cooling, programmable flux control and component-resolved detection within the same optical system.

\subsection{Spectrally engineered 626-nm MOT}

To realize the simultaneous dual-isotope MOT operation, we engineer the 626-nm spectrum before frequency conversion. In our apparatus, the 626-nm light is generated by sum-frequency generation of 1550-nm and 1050-nm seed lights. The isotope-resolved spectrum is programmed on the 1550-nm seed using a fiber Mach--Zehnder electro-optic modulator (MZM), and the imprinted frequency structure is transferred to generated the 626-nm light, which is then distributed to all MOT beams through common optical paths.

For each selected isotope pair, the MZM is operated in the carrier-suppressed double-sideband (CS-DSB) regime. The DC bias is set near the transmission minimum to suppress the carrier, while the RF drive amplitude is optimized to concentrate the optical power into the targeted first-order sidebands. The modulation frequency is set to half of the frequency separation between the selected 626-nm MOT transitions, including the relevant hyperfine shifts for the fermionic isotopes. 
After sum-frequency generation, the two first-order sidebands form the frequency components that address the corresponding isotopes simultaneously. One of these components is stabilized to the same ULE reference cavity through an auxiliary fiber EOM, yielding an optical linewidth of about 5~kHz. The auxiliary-EOM offset controls the absolute detuning of the two components together, whereas the MZM drive frequency determines their relative frequency separation.

We characterize the spectral purity of the engineered 626-nm light using the transmission spectrum of the ULE cavity~\cite{SM}. As shown in Fig.~\ref{fig:imple}(d), the residual carrier is suppressed by more than 30~dB, while the targeted first-order sidebands contain more than 96\% of the total optical power. The optical power is therefore concentrated predominantly in the two isotope-addressing components, with unwanted sidebands strongly suppressed. Because the two isotope-addressing frequency components are distributed through the same MOT beam paths, the two addressed isotopes experience closely matched optical trapping geometries, favoring strong spatial overlap. This implementation therefore enables simultaneous dual-isotope MOT and CMOT operation without additional isotope-specific cooling beam paths.

\section{Sequence-matched benchmarks with the $^{162}$Dy--$^{164}$Dy mixture}

\begin{figure*}[tb!]
	\includegraphics[width=1.9\columnwidth]{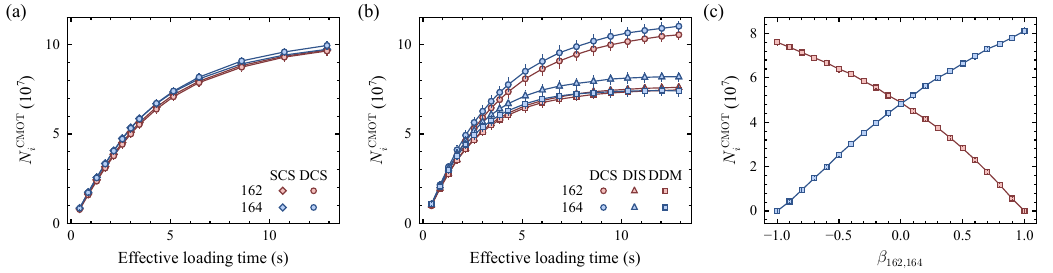}
	\caption{\textbf{Loading dynamics and population control of the \textsuperscript{162}Dy--\textsuperscript{164}Dy mixture.}
		(a) Single-isotope loading dynamics under SCS and DCS conditions for both isotopes, showing nearly identical loading behavior upon introducing the second 626-nm frequency component.
		(b) Loading dynamics under DCS, DIS, and DDM conditions, characterizing the effects of interrupted 421-nm addressing and full dual-isotope operation. Atom numbers are measured after the same subsequent CMOT sequence and plotted against the accumulated effective 421-nm loading time $T_i^{421}$.
		(c) $^{162}$Dy and $^{164}$Dy atom numbers after the CMOT stage as a function of the 421-nm loading bias $\beta_{162,164}$ after 20 switching cycles, demonstrating continuous control from strongly $^{162}$Dy-rich to $^{164}$Dy-rich mixtures.
	}
	\label{fig:loading}
\end{figure*}

We first benchmark the reconfigurable preparation sequence using the bosonic $^{162}$Dy--$^{164}$Dy mixture. Owing to their similar natural abundances, the two isotopes can be loaded with nearly balanced populations under comparable 421-nm dwell times, making this mixture a natural representative system for characterizing the complete preparation pathway. We first examine mixture formation and population control during the loading stage, and then follow the prepared sample through MOT holding, CMOT compression, and transfer into the ODT.

\subsection{Loading dynamics and population control}

To distinguish the effects of different operations on loading performance, we systematically benchmark dual-isotope mixture (DDM) together with three single-isotope controls (SCS, DCS, and DIS). Their three-letter labels specify, in order, the 626-nm spectrum, the 421-nm operation, and the atomic sample~\cite{SM}. SCS provides the single-isotope baseline, with a single 626-nm spectral component and continuous 421-nm addressing of the target isotope. DCS adds the second 626-nm spectral component, while DIS further introduces the 421-nm frequency-switching sequence, but with the light blocked during the companion-isotope dwell intervals to retain a single-isotope sample. In DDM, the 421-nm light remains on during both dwell intervals, producing a dual-isotope mixture confined by the two-frequency 626-nm MOT. The SCS--DCS and DCS--DIS comparisons therefore isolate the effects of the additional 626-nm spectral component and the interrupted 421-nm sequence, respectively. The DIS--DDM comparison provides the central benchmark for full dual-isotope operation, capturing the combined influence of companion-isotope 421-nm addressing and the presence of the second isotope. For isotope combinations exhibiting appreciable deviations, additional targeted diagnostics are implemented to assess the contribution of companion-isotope 421-nm light \cite{SM}.

Unless otherwise stated, data sets directly compared within the same panel were acquired in an interleaved sequence to minimize the effects of slow experimental drifts. For both MOT-loading and MOT-hold measurements, atom numbers were determined after the same subsequent CMOT sequence, irrespective of the preceding loading or holding duration.

As shown in Fig.~\ref{fig:loading}(a), the SCS and DCS loading curves nearly overlap for both isotopes, indicating little effect introduced by the additional 626-nm spectral component. In contrast, the DCS--DIS comparison in Fig.~\ref{fig:loading}(b) reveals an increasing difference with the accumulated effective 421-nm loading time $T_i^{421}$, where DIS exhibits a reduction of approximately $25\%$ near saturation. This behavior is consistent with the longer elapsed loading time and altered loading dynamics of the interrupted 421-nm sequence, providing a measure of its technical loading cost.

Under full dual-isotope operation, the DDM loading curve remains nearly identical to DIS for $^{162}$Dy, while $^{164}$Dy exhibits a modest additional reduction of approximately $10\%$. We quantify this comparison by the loading ratio $R_{i,N}^{\rm load}=N_{i,\rm DDM}/N_{i,\rm DIS}$. Since DDM differs from DIS in both the companion-isotope 421-nm addressing and the presence of the second isotope, this ratio captures their combined influence on loading rather than isolating a purely interspecies effect. 

As shown in Fig.~\ref{fig:loading}(c), the time-multiplexed 421-nm scheme also provides direct control over the mixture composition. By varying the 421-nm loading bias $\beta_{162,164}$, we tune the relative slow-atom fluxes delivered to the common 626-nm MOT and thereby control the populations after the CMOT stage. The accessible composition spans a broad range, from strongly $^{162}$Dy-rich to strongly $^{164}$Dy-rich samples, while the optical configuration and subsequent preparation sequence remain unchanged. For the $^{162}$Dy--$^{164}$Dy mixture, near-balanced populations are obtained for comparable dwell times and are used for the subsequent MOT-hold and CMOT measurements. This timing-based control largely decouples the prepared mixture composition from the natural-abundance ratio and provides a practical way of producing isotope mixtures with programmable population imbalance.

\begin{figure*}[tb!]
	\includegraphics[width=1.9\columnwidth]{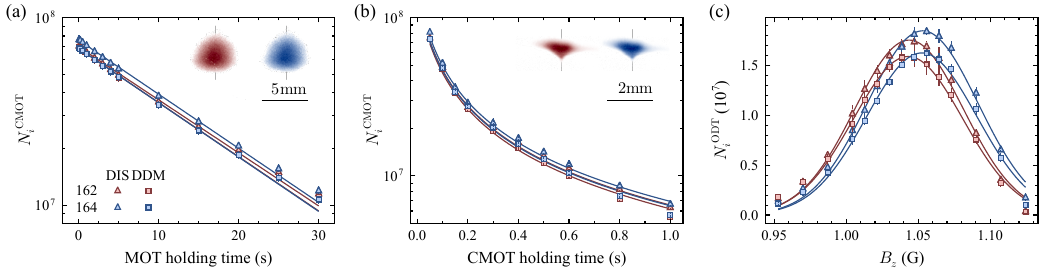}
	\caption{\textbf{Stability and ODT transfer of the $^{162}$Dy--$^{164}$Dy mixture.}
		(a) MOT-hold dynamics for DDM and DIS after the 421-nm slowing light is switched off. Atom numbers are plotted on a logarithmic scale, with solid curves showing exponential fits that yield lifetimes of around 15~s for all cases. Insets show component-resolved in-situ MOT images of $^{162}$Dy (red) and $^{164}$Dy (blue) under DDM conditions, demonstrating strong spatial overlap.
		(b) CMOT-hold dynamics for DDM and DIS, with solid curves showing fits to the effective two-body loss model. Insets show component-resolved in-situ CMOT images following the same convention as in (a), revealing closely matched cloud positions. 
		In both (a) and (b), the nearly identical DDM and DIS decay dynamics indicate no appreciable additional loss associated with dual-isotope operation.
		(c) ODT atom number versus the vertical offset B field $B_z$, which shifts the CMOT position relative to the ODT. Solid curves are corresponding Gaussian fits. For each isotope, DDM and DIS exhibit closely matched peak efficiencies, optimal loading positions, and widths. The loading optima of the two isotopes only differ slightly, allowing near-optimal transfer of both components under a common experimental condition.
		All measurements start from samples prepared with 20 switching cycles.		
	}
	\label{fig:stability}
\end{figure*}

\subsection{MOT stability and ODT transfer}

We next examine the effects introduced by the additional isotope during the subsequent preparation stages by comparing near-balanced $^{162}$Dy--$^{164}$Dy mixtures in DDM with the corresponding DIS single-isotope references. For the MOT- and CMOT-hold measurements, the 421-nm slowing light is switched off after loading, so that DDM and DIS share the same dual-frequency 626-nm environment and differ primarily by the presence of the second isotope. This comparison therefore directly assesses how the additional isotope affects trapped-sample stability and subsequent transfer into the ODT.

We first examine the stability of the trapped sample by holding the atoms in the MOT after the 421-nm slowing light is switched off. As shown in Fig.~\ref{fig:stability}(a), when plotted on a logarithmic atom-number scale, the decay curves remain approximately linear over the measured time range, consistent with predominantly one-body loss dynamics. 
Both isotopes exhibit nearly identical decay dynamics, and for each isotope the DDM curve closely follows the corresponding DIS reference. The fitted characteristic lifetimes are all around $15~\mathrm{s}$. We quantify this comparison using
$R_{i,\tau}^{\rm MOT}=\tau_{i, \rm DDM}/\tau_{i, \rm DIS}$,
which remains close to unity for both $^{162}$Dy and $^{164}$Dy. The nearly identical decay dynamics show that the presence of the second isotope does not introduce an appreciable additional loss channel in the MOT. The component-resolved in-situ images further show closely coincident cloud centers and comparable spatial extents, supporting substantial spatial overlap of the two isotopes.

The stability of the compressed sample is next examined by holding the atoms in the CMOT stage for a variable time before imaging. The atomic density in the CMOT is much higher than that in the MOT, and the atom number decays on a substantially shorter timescale, as shown in Fig.~\ref{fig:stability}(b), making this stage particularly sensitive to additional loss associated with the presence of the second isotope. 
The DDM components and their corresponding DIS references nevertheless show closely matched decay dynamics, apart from a slightly larger initial atom number for the DIS $^{164}$Dy reference. The decay is well described by the two-body-dominated form $N_i(t)=N_i(0)/[1+\kappa_i N_i(0)t]$, where $\kappa_i$ is used as an effective fitting parameter rather than as a microscopic two-body loss coefficient, with a larger $\kappa_i$ corresponding to faster atom loss. We characterize the additional-isotope effect by $R_{i,\kappa}^{\rm CMOT}=\kappa_{i,\mathrm{DIS}}/\kappa_{i,\mathrm{DDM}}$, such that $R_{i,\kappa}^{\rm CMOT}<1$ indicates enhanced loss in DDM relative to DIS. For the $^{162}$Dy--$^{164}$Dy mixture, $R_{i,\kappa}^{\rm CMOT}$ remains close to unity for both components. The component-resolved in-situ CMOT images shown in the inset of Fig.~\ref{fig:stability}(b) further reveal closely coincident cloud centers and comparable spatial distributions, supporting substantial spatial overlap of the two isotopes after compression.

We finally benchmark the transfer of the compressed clouds into the acousto-optic-deflector-scanned single-beam ODT, as shown in Fig.~\ref{fig:stability}(c). We vary the vertical offset B field $B_z$, which shifts the CMOT position relative to the ODT during loading, while keeping the CMOT parameters, ODT settings, and loading sequence unchanged. The measurements are fitted with Gaussian functions to determine the optimal loading conditions. For each component, we define the transfer efficiency as $\eta_i=N_{i}^{\rm ODT}/N_{i}^{\rm CMOT}$, with $N_{i}^{\rm CMOT}$ the corresponding CMOT atom numbers, the fitted peaks yield the peak transfer efficiencies $\eta_{i}^{\rm max}$.

For a given isotope, the DDM and DIS loading curves nearly overlap, with closely matched fitted peak positions and widths. The small differences in the peak atom numbers mainly arise from the slightly different initial CMOT atom numbers, while the corresponding peak transfer efficiencies are all close to $22\%$. This shows that introducing the second isotope does not appreciably modify the spatial ODT-loading response. The optimal loading positions of $^{162}$Dy and $^{164}$Dy exhibit a small relative offset, but their loading windows strongly overlap. Together with the component-resolved in-situ MOT and CMOT images, the loading-position measurements provide an independent check that the two isotope clouds remain well spatially overlapped prior to ODT loading. A common operating point therefore allows both components to be transferred close to their respective peak efficiencies, requiring little compromise between the two isotopes.

\section{Reconfigurability across isotope pairs}

\begin{figure*}[tb!]
	\includegraphics[width=1.8\columnwidth]{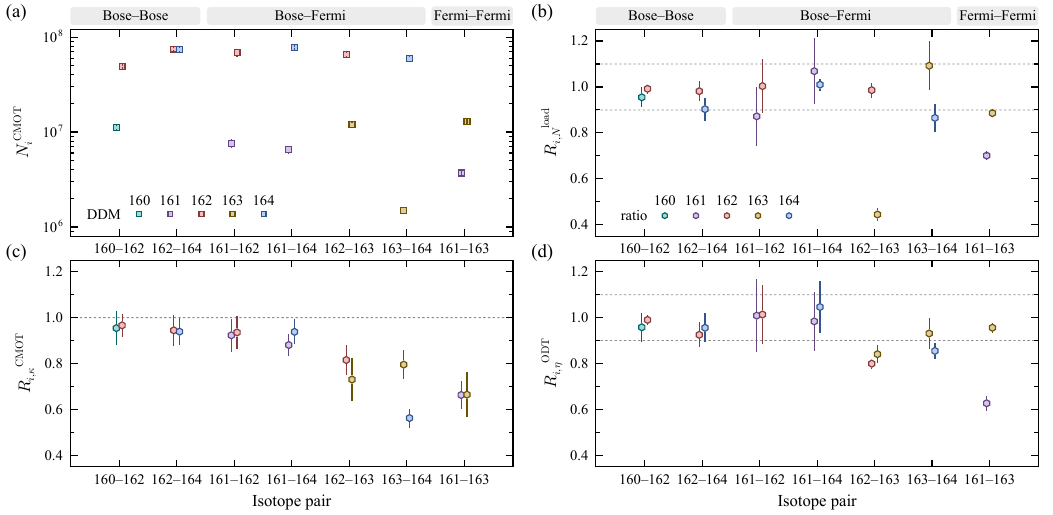}
	\caption{\textbf{Reconfigurability across isotope combinations.}
		(a) Component-resolved atom numbers after the CMOT stage for Bose--Bose, Bose--Fermi, and Fermi--Fermi isotope mixtures, measured with nearly equal 421-nm dwell times. MOT-loading conditions are optimized separately for each isotope pair. 
		(b) Loading ratio $ R_{i,N}^{\rm load}=N_{i,\rm DDM}/N_{i,\rm DIS}$, comparing dual-isotope loading with the corresponding interrupted single-isotope reference. 
		(c) CMOT stability ratio $ R_{i,\kappa}^{\rm CMOT}=\kappa_{i,\rm DIS}/\kappa_{i,\rm DDM}$ obtained from CMOT-hold measurements. Values below unity correspond to faster decay in DDM. 
		(d) ODT-transfer ratio $R_{i,\eta}^{\rm ODT}=\eta_{i,\rm DDM}^{\rm max}/\eta_{i,\rm DIS}^{\rm max}$, where $\eta_{i,\rm DDM}^{\rm max}$ and $\eta_{i,\rm DIS}^{\rm max}$ are the peak transfer efficiencies obtained from Gaussian fits to the corresponding ODT-loading curves. Ratios near unity indicate that dual-isotope operation introduces little additional reduction in the maximum transfer efficiency relative to the corresponding DIS reference. 
		Panels (a) and (b) are obtained with 30 switching cycles per shot, whereas (c) and (d) use samples prepared with 20 switching cycles.
		Ratios near unity in (b)--(d) indicate near identical DDM and DIS performance for the corresponding preparation stage. %Colors denote isotope identity, while marker shapes distinguish the preparation protocols.
	}
	\label{fig:isotopes}
\end{figure*}

Building on the $^{162}$Dy--$^{164}$Dy benchmark, we extend the same multiplexed preparation sequence to additional isotope pairs spanning Bose--Bose, Bose--Fermi, and Fermi--Fermi mixtures, as shown in Fig.~\ref{fig:isotopes}. The cooling-beam geometry and optical paths remain unchanged, while isotope-pair reconfiguration is implemented by updating the RF-defined 421-nm switching frequencies, 626-nm sideband separation, and locking offsets. Because the different isotopes generally require different MOT-loading conditions, the loading parameters for each pair are chosen primarily to optimize the more weakly loaded component, while the 421-nm dwell-time bias $\beta_{i,j}$ is used to adjust the relative populations. Near-balanced samples are prepared specifically for the MOT- and CMOT-hold measurements, where comparable component populations provide a clearer benchmark of the stability of the two-component sample.

For mixtures containing fermionic isotopes, both the 421-nm and 626-nm cooling light are primarily tuned to address the $F=21/2\rightarrow F'=23/2$ cycling transition. Notably, the $F=21/2$ level is the lowest hyperfine level for $^{161}$Dy but not for $^{163}$Dy. Because of its broader linewidth, the 421-nm transition can also accidentally couple to nearby hyperfine transitions, whereas the 626-nm transition provides more spectrally selective addressing. More efficient MOT loading of $^{163}$Dy than $^{161}$Dy has nevertheless been observed and attributed to favorable optical pumping~\cite{Lu2012}. 

Figure~\ref{fig:isotopes}(a) summarizes the component-resolved atom numbers obtained after the CMOT stage for the different isotope combinations under nearly equal 421-nm dwell times. Two-component samples are obtained across the Bose--Bose, Bose--Fermi, and Fermi--Fermi mixtures, although the absolute atom numbers vary substantially between isotope pairs because of differences in natural abundance and the corresponding optimized MOT conditions. These absolute numbers therefore characterize the accessible sample sizes under comparable dwell-time allocation rather than providing a direct comparison of intrinsic preparation performance between different isotope pairs. To evaluate the additional change associated with dual-isotope operation, we use the loading ratio $ R_{i,N}^{\rm load}=N_{i,\rm DDM}/N_{i,\rm DIS}$ shown in Fig.~\ref{fig:isotopes}(b). For most isotope combinations, $ R_{i,N}^{\rm load}$ remains close to unity, indicating that the atom numbers obtained in DDM remain close to the corresponding interrupted single-isotope references even with the companion-isotope 421-nm addressing and the second isotope are introduced.

The larger deviations in $ R_{i,N}^{\rm load}$ are concentrated in several $^{163}$Dy-containing combinations and depend strongly on the selected isotope pair. For $^{162}$Dy--$^{163}$Dy, auxiliary single-isotope measurements show that the reduced $^{163}$Dy loading originates predominantly from the companion-isotope 421-nm Zeeman-slower light~\cite{SM}. This sensitivity is consistent with the $^{162}$Dy slowing frequency lying close to the zero-field $^{163}$Dy transition. By contrast, for $^{163}$Dy--$^{164}$Dy, the DCS--SCS comparison reveals a pronounced sensitivity of the $^{163}$Dy loading to the additional 626-nm spectral component. The $^{161}$Dy--$^{163}$Dy Fermi--Fermi mixture provides another representative case, where the additional 626-nm component produces only a modest change, while the companion-isotope 421-nm addressing accounts for most of the larger reduction observed in DDM~\cite{SM}. These results show that the observed loading reductions arise from isotope-pair-specific sensitivities to different parts of the preparation sequence rather than from a generic degradation associated with dual-isotope operation.

To distinguish the loading-stage effects from the subsequent stability of the trapped mixtures, we compare nearly number-balanced DDM and DIS samples using the same hold measurements as in Fig.~\ref{fig:stability}(a) and (b), including both the low-density MOT and the higher-density CMOT conditions, and with all 421-nm light switched off during the hold. In the MOT, the lifetime ratio $ R_{i,\tau}^{\rm MOT}=\tau_{i,\rm DDM}/\tau_{i,\rm DIS}$ remains close to unity for both components across the achieved isotope combinations, including several $^{163}$Dy-containing mixtures that show more pronounced deviations during loading~\cite{SM}. Thus, the reduced atom numbers observed during loading for some isotope pairs are not accompanied by faster MOT decay, indicating that the presence of the second isotope introduces little additional loss in the trapped MOT. In the much higher-density CMOT, however, stronger isotope-pair dependence emerges. As shown in Fig.~\ref{fig:isotopes}(c), the CMOT stability ratio $ R_{i,\kappa}^{\rm CMOT}=\kappa_{i,\rm DIS}/\kappa_{i,\rm DDM}$ remains close to unity for most isotope combinations, indicating comparable decay rates in DDM and DIS and little additional density-dependent loss associated with the second isotope. Larger deviations occur among the $^{163}$Dy-containing mixtures. For $^{163}$Dy--$^{164}$Dy, part of this sensitivity is already present in the single-isotope reference due to the additional 626-nm spectral component~\cite{SM}. The $^{161}$Dy--$^{163}$Dy mixture exhibits one of the most pronounced deviations. Since $\kappa_i$ is used only as a comparative decay parameter, these deviations characterize high-density CMOT stability rather than microscopic interspecies loss coefficients. Together, the MOT- and CMOT-hold measurements show that the second isotope introduces little additional loss in the low-density MOT for almost all combinations, while stronger isotope-pair dependence emerges after compression.

We finally compare ODT transfer across the different isotope combinations. We define the ratio $R_{i,\eta}^{\rm ODT}=\eta_{i,\rm DDM}^{\rm max}/\eta_{i,\rm DIS}^{\rm max}$, where $\eta_{i,\rm DDM}^{\rm max}$ and $\eta_{i,\rm DIS}^{\rm max}$ are the peak transfer efficiencies obtained from Gaussian fits to the corresponding loading curves. This comparison characterizes the additional transfer penalty associated with dual-isotope operation comparing to the single-isotope reference. As shown in Fig.~\ref{fig:isotopes}(d), $R_{i,\eta}^{\rm ODT}$ remains close to unity for most components, showing that their peak ODT-transfer efficiencies are largely preserved under dual-isotope operation. Appreciable reductions occur in certain cases, most notably the $^{161}$Dy component of the $^{161}$Dy--$^{163}$Dy mixture.

\section{Conclusions and outlook}

In this work, we have established a reconfigurable preparation scheme for bosonic and fermionic dysprosium isotope mixtures while retaining the simplicity of a single-isotope experimental setup. This reconfigurability extends across Bose--Bose, Bose--Fermi, and Fermi--Fermi combinations, with flexible isotope-pair selection and tunable population balance. Systematic benchmarks demonstrate that most components retain nearly single-isotope-like performance throughout the preparation sequence, with appreciable deviations limited to certain isotope combinations and stages. Targeted controls identify several loading-stage deviations to pair-dependent optical sensitivities. While the low-density MOT exhibits little additional loss under dual-isotope operation, enhanced decay emerges for certain combinations at higher CMOT densities. Further component-dependent differences are observed during ODT transfer. These results demonstrate the robustness of the common preparation pathway while identifying the specific stages where further optimization is most relevant for extending different isotope combinations toward simultaneous quantum degeneracy.

The significance of this platform extends beyond the preparation of different isotope combinations. Dysprosium isotopes combine closely matched single-particle properties with component-selective optical addressability, allowing different mixtures to be prepared under similar trapping conditions. The availability of both bosonic and fermionic isotopes further makes quantum statistics selectable through isotope-pair choice, while their strong long-range anisotropic dipolar interactions provide access to a distinctive regime of multicomponent quantum matter. Unlike many heteronuclear mixtures, isotope-pair selection and population control can therefore be achieved without introducing substantial changes in mass, polarizability, or the underlying trapping environment. This combination offers a particularly favorable setting for investigating the effects of composition, intercomponent coupling, and quantum statistics while minimizing the difference of underlying single-particle environments.

Extending these preparations toward simultaneous quantum degeneracy, together with tunable intra- and inter-species interactions, would open new opportunities to explore the interplay of quantum statistics, composition, and strong dipolar coupling. Continuous population control could connect balanced mixtures to strongly imbalanced and impurity regimes, enabling studies of dipolar impurities, mediated interactions, and coupled density--composition dynamics. Particularly rich physics may emerge near interaction-driven self-organization, where intercomponent coupling and population imbalance can greatly alter roton instabilities and the formation of quantum droplets and supersolids, giving rise to collective ordering beyond the single-component regime. By enabling comparisons across bosonic and fermionic mixtures under closely matched trapping conditions, this platform offers a distinctive route to uncovering how quantum statistics, composition, and intercomponent coupling reshape collective behavior and emergent order in strongly dipolar quantum matter.

\begin{acknowledgements}
	
	This work is supported by the National Key R$\&$D Program of China (No.~2022YFA1405800), Innovation Program for Quantum Science and Technology (No.~2024ZD0300600), National Natural Science Foundation of China (No.~12474263), Guangdong Basic and Applied Basic Research Foundation (No.~2023B0303000011), Guangdong Provincial Quantum Science Strategic Initiative (No.~GDZX2304006, GDZX2303002, GDZX2403002), Shenzhen Science and Technology Program (No.~KQTD20240729102026004), Stable Support Plan Program of Shenzhen Natural Science Fund (No.~20220815001356001), and Guangdong Provincial Key Laboratory of Advanced Thermoelectric Materials and Device Physics (No.~2024B1212010001).

\end{acknowledgements}

%\bibliographystyle{apsrev4-2}
%\bibliography{refs_all_clean} 
\input{Mixtureref.bbl}

\setcounter{figure}{0}
\renewcommand{\thefigure}{S\arabic{figure}}

\setcounter{table}{0}
\renewcommand{\thetable}{S\arabic{table}}

\setcounter{equation}{0}
\renewcommand{\theequation}{S\arabic{equation}}

\clearpage
\onecolumngrid

\section*{Supplemental Material}

\subsection{Experimental setup and sequence}

The overall experimental sequence retains the same structure as the single-isotope preparation, with dual-isotope operation implemented primarily through time-multiplexed frequency switching of the 421-nm light and spectral engineering of the 626-nm MOT light. A typical sequence consists of MOT loading, compression, and transfer into an optical dipole trap. During MOT loading, the 421-nm light provides transverse cooling and Zeeman slowing and is time multiplexed between the two isotopes. Each switching cycle lasts 1~s. For equal dwell times, the 421-nm light addresses each isotope for about 430~ms per cycle, while two frequency-switching intervals with the 421-nm lights actively extinguished of about 70~ms each give a total dead time of about 140~ms. This dead time is kept fixed when the dwell-time bias is varied, so that the total available addressing time of the two isotopes remains approximately 860~ms in each cycle. After \(n\) switching cycles, the accumulated effective loading time for isotope \(i\) is therefore \(T_i^{421}=n\times t_i\), where \(t_i\) denotes its dwell time within one cycle. Unless otherwise specified, the MOT-hold, CMOT-hold, population-control, and ODT-transfer measurements are performed with 20 switching cycles of MOT loading for each realization, for which the MOT populations are already close to saturation. The cross-isotope CMOT populations and loading ratios summarized in Fig.~5(a) and (b) are evaluated with 30 switching cycles of MOT loading.

During the loading stage, the atoms are captured in a five-beam 626-nm MOT. The 626-nm light is spectrally broadened by approximately \(1~\mathrm{MHz}\) using a low-frequency resonant EOM to increase the capture velocity and is operated at a red detuning of approximately \(80~\Gamma\), where \(\Gamma=2\pi\times136~\mathrm{kHz}\) is the natural linewidth of the 626-nm transition. The magnetic-field gradient and bias fields are adjusted accordingly for this far-detuned loading configuration. Under these conditions, the MOT is dilute and its equilibrium position lies well below the center of the apparatus, as illustrated in the inset of Fig.~1(b). At the end of MOT loading, the 421-nm cooling and slowing light and the spectral broadening of the 626-nm light are switched off. The cloud is then compressed through two consecutive ramps of 95~ms and 40~ms, during which the magnetic-field gradient and bias fields, together with the 626-nm frequency and optical power, are varied simultaneously. This is followed by a 55-ms hold at the final CMOT parameters. The final CMOT conditions are optimized for each isotope pair, although the resulting operating points remain similar across the different combinations. The MOT and especially the CMOT parameters are chosen to optimize subsequent loading into the ODT rather than to maximize the atom number after compression. During the compression sequence, the cloud is therefore not only compressed but also displaced upward toward the center of the apparatus, where the ODT is located.

Unless otherwise stated, the CMOT atom number is measured after this standard compression sequence. For the MOT-hold measurements, the 421-nm light is switched off immediately after MOT loading, while the 626-nm spectral broadening, frequency, optical power, magnetic-field gradient, and bias fields are maintained at their loading values throughout a variable hold time. The standard CMOT sequence is then applied before detection. For the CMOT-hold measurements, the variable hold time replaces the standard 55-ms hold at the final CMOT parameters. The two measurements therefore probe the stability of the sample under the corresponding low-density MOT and higher-density CMOT conditions while otherwise retaining the same preparation and detection sequence.

The ODT is formed from a single-frequency 1064-nm fiber laser. A single Gaussian beam with a waist radius of approximately \(27~\mu\mathrm{m}\) is scanned horizontally with an acousto-optic deflector at approximately 70~kHz. The AOD is driven with an inverse-cosine frequency scan, producing a time-averaged flattened trapping potential with a horizontal-to-vertical aspect ratio of about 10. The optical power is maintained at approximately 33~W during loading. The ODT is switched on at the beginning of the CMOT compression, when the atomic cloud is still well below the trapping region. As the CMOT sequence proceeds, the upward displacement of the cloud brings it into spatial overlap with the ODT. At the end of the CMOT sequence, the 626-nm light is switched off and the atoms are held in the ODT for approximately 60~ms before release and detection.

Atom numbers, temperatures, and in-situ MOT and CMOT density distributions are obtained using side-view absorption imaging on the 421-nm transition. Only one isotope is imaged in each experimental realization, with the two components measured in interleaved runs by switching the imaging frequency to the corresponding isotope transition. The time of flight is chosen according to the observable. In-situ MOT and CMOT images are acquired without expansion, whereas CMOT atom numbers are typically measured after 30-ms TOF and ODT-loaded samples after 7~ms. Because the loading-stage MOT is located well below the center of the apparatus, its in-situ imaging is performed at approximately one-quarter of the standard magnification to provide a sufficiently large field of view. All other side-view imaging is performed at approximately \(1{:}1\) magnification.

\subsection{Spectral performance of the MZM modulation on the 626-nm light}

\begin{figure}[tb!]
	\includegraphics[width=0.9\columnwidth]{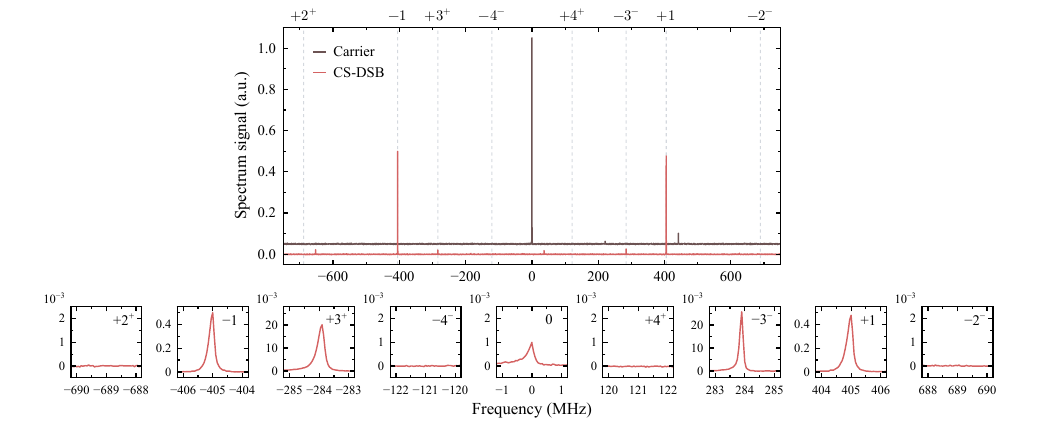}
	\caption{\textbf{Transmission spectral signal under the Carrier and CS-DSB settings.} The full spectrum within one FSR around the carrier acquired by scanning the laser frequency. Sideband orders of fundamental cavity modes (dashed lines) are patched with fine-scanned data whose zoomed-in views are listed below, each labeled with the signed integer order and the FSR index as superscript. Other tiny peaks arise from the higher TEM modes of the cavity due to imperfect coupling. The Carrier spectrum is given a small vertical offset for visual clarity. }
	\label{fig:spectra626}
\end{figure}
We map out the MZM-modulated 626-nm spectrum by observing the ULE cavity's transmission signal while scanning the laser frequency under two settings. Under the CS-DSB setting, the MZM is driven by an $\Omega = 405$~MHz RF signal, and the DC input is locked to the minimal-transmission null point to suppress the carrier component. This generates a series of sidebands of different heights, each labeled with signed order $m$ and has a frequency offset of $m\Omega$ from the carrier. For comparison, under the Carrier setting, the MZM has no RF input, and the DC input is locked to the maximal-transmission peak point to maintain a stable power. These same settings are used to generate the double- or single- frequency 626-nm spectrum in the measurements. Fig.~\ref{fig:spectra626} shows the measured spectral signal under each setting from the combination of a single full-range scan and the averaged fine-scan signals around each sideband order.

The frequency scanning is conducted by ramping the PZT bias to the 1050-nm seed (as shown in Fig.~2(b)). The cavity transmission spectrum is folded within one FSR, and we choose the carrier frequency as the center. A higher-order sideband would fall into the FSR indexed by $p$, but appears on the spectrum as being folded to $m\Omega - p \cdot \text{FSR}$, hence labeled as $m^p$ in Fig. \ref{fig:spectra626} ($p = \pm$ abbreviated for $\pm 1$). A peak at such sideband order is identified using a temporary reference peak generated by the fiber-EOM driven at the corresponding frequency $m\Omega$, which results in an overlap. The reference frequencies have an uncertainty of about 100~kHz, providing a set of calibrated frequency markers across the scanned spectral range.

For the full-range scan, these calibrated markers are used to convert the PZT-scan voltage to optical frequency. Measurements acquired over several scans are combined, and the frequency conversion is obtained from a cubic-spline fit to the calibrated markers. Repeated measurements give a frequency uncertainty below 10~MHz over the full spectrum. For the fine-scan spectra in Fig.~S1, the frequency axis is calibrated locally and linearly around each sideband using the corresponding fiber-EOM reference.

The spectrum shows that the first-order sidebands under the CS-DSB setting take up more than 96\% of the total power, providing addressings to the dual-isotope atoms, while the carrier is suppressed by 30~dB, proving that the redistribution of optical power into the first-order sidebands is efficient. 

\subsection{Loading-control hierarchy and systematic benchmarks}

To systematically identify the loading changes introduced when extending the single-isotope preparation to dual-isotope operation, we employ the matched protocols summarized in Table~\ref{tab:protocols}. The three-letter labels specify, respectively, the 626-nm spectrum, the 421-nm operation, and the atomic sample. The first letter denotes a single- (S) or dual-isotope (D) 626-nm spectrum, the second denotes continuous (C), interrupted (I), or dual-isotope (D) 421-nm addressing, and the third denotes a single-isotope sample (S) or an isotope mixture (M). Together, these protocols form a sequence of controlled comparisons connecting the single-isotope reference to the full dual-isotope loading condition.

\begin{table}[t]
	\caption{\textbf{Experimental protocols used for the loading comparisons.}
		For the 626-nm spectrum, the single- or dual- frequency settings follow the settings of Carrier and CS-DSB. For the 421-nm operation, in the interrupted 421-nm sequence, the laser follows the same frequency-switching sequence as in dual-isotope operation, but the light is blocked during the dwell interval assigned to the companion isotope. In the dual-addressed sequence, the 421-nm light is applied during both isotope-specific dwell intervals. A mixture sample is prepared only when both 626-nm spectrum and 421-nm operation enable dual-isotope addressing.}
	\label{tab:protocols}
	\centering
	\setlength{\tabcolsep}{4pt}
	\begin{tabular}{c c c c}
		\toprule  
		Protocol & 626-nm spectrum & 421-nm operation & Sample \\
		\midrule
		SCS & Single & Continuous & Single isotope \\
		DCS & Dual & Continuous & Single isotope \\
		SIS & Single & Interrupted & Single isotope \\
		DIS & Dual & Interrupted & Single isotope \\
		SDS & Single & Dual addressed & Single isotope \\
		DDM & Dual & Dual addressed & Mixture \\
		\bottomrule  
	\end{tabular}
\end{table}

\begin{figure}[tb!]
	\includegraphics[width=0.9\columnwidth]{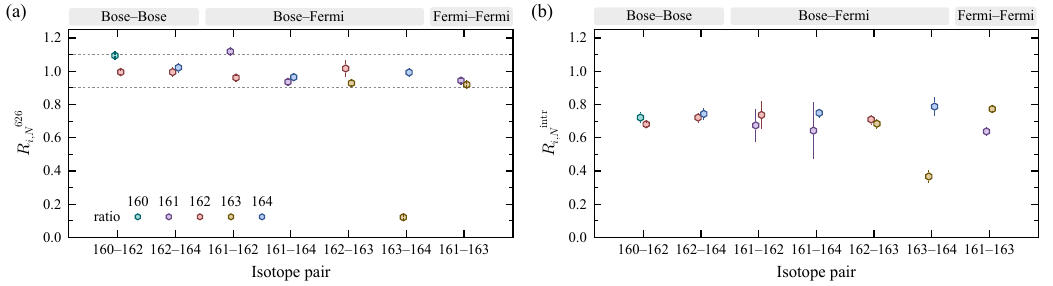}
	\caption{\textbf{Systematic loading controls across dysprosium isotope combinations.}
		(a) Ratio $R_{i,N}^{626}=N_{i,\rm DCS}/N_{i,\rm SCS}$, comparing DCS with the corresponding SCS reference and characterizing the influence of the dual-frequency 626-nm MOT spectrum on target-isotope loading.
		(b) Ratio $R_{i,N}^{\rm intr}=N_{i,\rm DIS}/N_{i,\rm DCS}$, comparing DIS with DCS and characterizing the loading reduction associated with the interrupted 421-nm sequence. The interrupted sequence is operated under the the balanced 421-nm dwell-time setting, $t_i=t_j$. All data are obtained with effective 30 switching cycles.}
	\label{fig:figs1}
\end{figure}

These protocols are chosen to separate the different effects introduced along the loading sequence. The SCS--DCS comparison characterizes the sensitivity of target-isotope loading to replacing the single-frequency 626-nm MOT spectrum by the dual-frequency configuration while retaining continuous 421-nm addressing. The DCS--DIS comparison characterizes the loading change associated with the interrupted 421-nm sequence, including the reduced target-isotope addressing duty cycle and the switching dead time, while keeping the dual-frequency 626-nm configuration unchanged. The DIS--DDM comparison, presented in Fig.~5(b) of the main text, benchmarks full dual-isotope loading against the corresponding interrupted single-isotope reference. In this comparison, the companion-isotope 421-nm light is restored and the second isotope is simultaneously introduced, so the two contributions are not directly separated. For components showing a pronounced reduction in the DIS--DDM comparison, we therefore perform additional SIS--SDS measurements as targeted controls. SIS and SDS have the same single-frequency 626-nm configuration and the same 421-nm switching timing but differ in whether the companion-isotope 421-nm light is blocked or applied during its assigned dwell interval, thereby directly testing the influence of this light without introducing a second trapped isotope.

The first two comparisons are performed systematically across the isotope combinations studied and are summarized in Fig.~\ref{fig:figs1}. Figure~\ref{fig:figs1}(a) shows $R_{i,N}^{626}=N_{i,\rm DCS}/N_{i,\rm SCS}$, which characterizes the effect of introducing the dual-frequency 626-nm MOT spectrum. SCS and DCS retain continuous 421-nm addressing of the target isotope, with all other MOT parameters kept unchanged. The total 626-nm power in DCS is set to around twice that in SCS, such that the power carried by the target-isotope spectrum remains approximately unchanged when the companion-isotope component is added. A value of $R_{i,N}^{626}$ close to unity therefore indicates little influence of the dual-frequency 626-nm configuration on target-isotope loading. As shown in Fig.~\ref{fig:figs1}(a), this is the case for most isotope combinations. A clear exception is the $^{163}$Dy component of the $^{163}$Dy--$^{164}$Dy combination, for which $R_{i,N}^{626}$ is around 0.1, consistent with the pronounced suppression of the absolute $^{163}$Dy atom number for this combination in Fig.~\ref{fig:isotopes}(a). This pronounced reduction indicates that the degraded $^{163}$Dy loading in this combination is predominantly associated with the introduction of the additional 626-nm spectral component.

Figure~\ref{fig:figs1}(b) characterizes the loading change introduced by the interrupted 421-nm sequence through $R_{i,N}^{\rm intr}=N_{i,\rm DIS}/N_{i,\rm DCS}$, which quantifies the loading penalty associated with interrupted target-isotope addressing. DCS and DIS employ the same dual-frequency 626-nm configuration and otherwise identical MOT parameters, but differ in the 421-nm operation. In DIS, the 421-nm laser follows the same frequency-switching sequence as in dual-isotope loading, while the light is blocked during the dwell interval assigned to the companion isotope. As shown in Fig.~\ref{fig:figs1}(b), $R_{i,N}^{\rm intr}$ is around 0.7 for nearly all isotope combinations. The similar values across different isotope pairs indicates that this reduction is primarily a common technical consequence of the time-multiplexed 421-nm loading sequence rather than an isotope-pair-specific effect. The main exception is the $^{163}$Dy component of the $^{163}$Dy--$^{164}$Dy combination, which also exhibits the pronounced sensitivity to the dual-frequency 626-nm configuration identified in Fig.~\ref{fig:figs1}(a).

Figure~\ref{fig:isotopes}(b) of the main text characterizes the additional loading change upon going from the interrupted single-isotope reference to full dual-isotope operation through $R_{i,N}^{\rm load}=N_{i,\rm DDM}/N_{i,\rm DIS}$. DDM and DIS employ the same dual-frequency 626-nm spectrum, 421-nm switching timing, target-isotope dwell time, number of switching cycles, and otherwise identical MOT parameters. They differ in that, in DDM, the 421-nm light is applied during the companion-isotope dwell interval and the second isotope is simultaneously trapped. The ratio $R_{i,N}^{\rm load}$ therefore contains the combined influence of the companion-isotope 421-nm light and the presence of the second isotope and does not by itself separate these two contributions. As shown in Fig.~5(b), $R_{i,N}^{\rm load}$ remains close to unity for most isotope components, whereas the more pronounced reductions occur predominantly for $^{163}$Dy-containing mixtures.

\subsection{Pair-dependent loading deviations and targeted diagnostics}

\begin{figure}[tb!]
	\includegraphics[width=0.9\columnwidth]{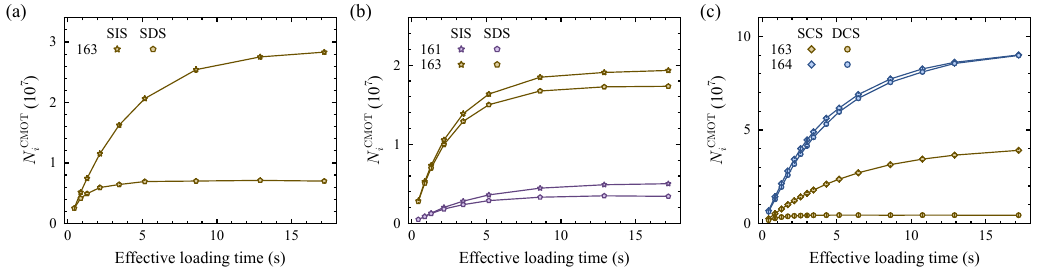}
	\caption{\textbf{Targeted diagnostics of pair-dependent loading deviations.}
		(a) SIS and SDS loading curves for the $^{163}$Dy component of the $^{162}$Dy--$^{163}$Dy combination. The companion-isotope 421-nm light greatly reduces the $^{163}$Dy loading performance.
		% to $R_{163,N}^{421}\simeq0.4$, close to the corresponding ratio obtained under full dual-isotope loading in Fig.~5(b).
		Only the $^{163}$Dy component is measured here, since $R_{162,N}^{\rm load}$ remains close to unity.
		(b) SIS and SDS loading curves for the $^{161}$Dy--$^{163}$Dy combination. The corresponding loading ratios are approximately $R_{161,N}^{421}\simeq0.7$ and $R_{163,N}^{421}\simeq0.9$, consistent with the reductions observed under full dual-isotope loading in Fig.~5(b).
		(c) Detailed SCS and DCS loading curves for the $^{163}$Dy--$^{164}$Dy combination underlying the corresponding $R_{i,N}^{626}$ values in Fig.~\ref{fig:figs1}(a).
		}
	\label{fig:figs2}
\end{figure}

The systematic loading measurements above show that the pronounced deviations are concentrated in $^{163}$Dy-containing mixtures. At the same time, as discussed below, the corresponding MOT-hold measurements show no obvious additional loss associated with the presence of the second isotope in the low-density MOT. We therefore examine the optical origin of these deviations using targeted control measurements.
The reduced $^{163}$Dy loading in the $^{163}$Dy--$^{164}$Dy combination has already been identified above as arising predominantly from the dual-frequency 626-nm configuration, with the corresponding detailed SCS--DCS loading curves shown in Fig.~\ref{fig:figs2}(c). We therefore focus here on the remaining two $^{163}$Dy-containing mixtures that show clear deviations in Fig.~5(b), namely $^{162}$Dy--$^{163}$Dy and $^{161}$Dy--$^{163}$Dy. 

For these two combinations, the targeted SIS--SDS controls are shown in Figs.~\ref{fig:figs2}(a) and \ref{fig:figs2}(b). SIS and SDS use the same single-frequency 626-nm configuration and the same 421-nm switching timing, but differ in whether the companion-isotope 421-nm light is blocked or applied during its assigned dwell interval. We characterize this effect by $R_{i,N}^{421}=N_{i,\rm SDS}/N_{i,\rm SIS}$, which directly probes the influence of the companion-isotope 421-nm light.

For the $^{162}$Dy--$^{163}$Dy combination, only the $^{163}$Dy component is measured in this comparison, since the $^{162}$Dy component ratio retains close to unity in Fig.~5(b). As shown in Fig.~\ref{fig:figs2}(a), the companion-isotope 421-nm light already greatly suppress the $^{163}$Dy loading performance, indicating that this optical effect is likely the dominant contribution. 
This enhanced sensitivity is consistent with the proximity of the $^{162}$Dy slowing frequency to the zero-field optical resonance of $^{163}$Dy.

For the $^{161}$Dy--$^{163}$Dy combination, both components are shown in Fig.~\ref{fig:figs2}(b). The corresponding ratios are approximately $R_{161,N}^{421}\simeq0.7$ and $R_{163,N}^{421}\simeq0.9$, consistent with the relative reductions observed under full dual-isotope loading in Fig.~5(b). Here the larger relative reduction occurs for $^{161}$Dy rather than $^{163}$Dy, showing that the sensitivity to the companion-isotope 421-nm light is component dependent. For both isotope combinations, the close correspondence between $R_{i,N}^{421}$ and the corresponding $R_{i,N}^{\rm load}$ shows that the principal loading deviations observed between DIS and DDM can be largely reproduced by introducing the companion-isotope 421-nm light alone. 

Figure~\ref{fig:figs1} and \ref{fig:figs2}, together with Fig.~5(b) of the main text, provide a stage-resolved interpretation of the isotope-dependent loading performance. The interrupted 421-nm sequence introduces a nearly common loading penalty of approximately 30\% across most isotope combinations, reflecting the reduced addressing duty cycle and finite switching dead time. Beyond this common technical cost, the larger pair-dependent reductions are concentrated in $^{163}$Dy-containing mixtures and arise predominantly from specific optical sensitivities. The dual-frequency 626-nm configuration is the principal limitation for $^{163}$Dy in the $^{163}$Dy--$^{164}$Dy combination, whereas the companion-isotope 421-nm light accounts for most of the additional reduction in the $^{162}$Dy--$^{163}$Dy and $^{161}$Dy--$^{163}$Dy combinations. Together with the MOT-hold measurements discussed below, these controls separate the common technical cost of time-multiplexed loading from pair--specific optical sensitivities and show that the observed loading degradation is predominantly associated with the optical preparation rather than with a generic instability of the isotope mixtures.

\subsection{MOT and CMOT decay dynamics}

\begin{figure}[tb!]
	\includegraphics[width=0.9\columnwidth]{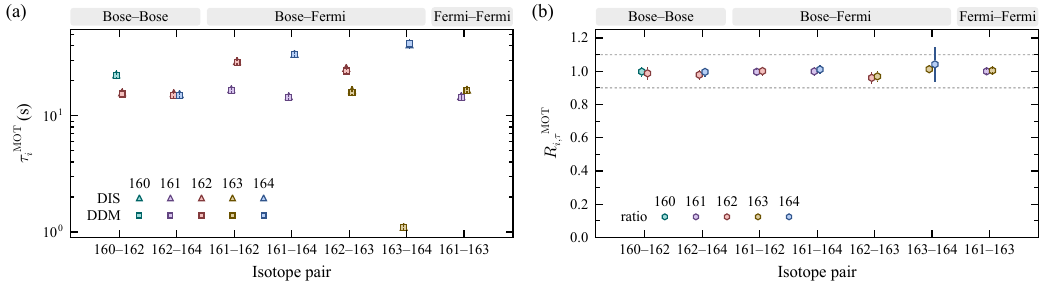}
	\caption{\textbf{MOT stability across dysprosium isotope combinations.}
		(a) MOT lifetimes $\tau_i$ for DDM and the corresponding DIS reference under near $n$-balanced condition. Except for the $^{163}$Dy component of the $^{163}$Dy--$^{164}$Dy combination, whose lifetime is strongly reduced by its sensitivity to the dual-frequency 626-nm configuration, all measured lifetimes exceed 10~s.
		(b) Corresponding lifetime ratio $R_{i,\tau}^{\rm MOT}=\tau_{i,\rm DDM}/\tau_{i,\rm DIS}$. The ratios remain very close to unity for all isotope pairs, including the short-lived $^{163}$Dy component of the $^{163}$Dy--$^{164}$Dy combination, indicating no appreciable additional loss associated with the presence of the second isotope in the low-density MOT.}
	\label{fig:figs3a}
\end{figure}

To examine whether the presence of the second isotope introduces additional loss after loading, we compare the hold dynamics of DDM with the corresponding DIS reference in both the low-density MOT and the compressed MOT. For all isotope pairs measured, the MOT atom number exhibits an approximately exponential decay over the investigated hold time, consistent with predominantly one-body loss under the low-density MOT conditions. We therefore fit the data with $N_i(t)=N_i(0)e^{-t/\tau_i}$.
The measurements are performed under near $n$-balanced conditions. As shown in Fig.~\ref{fig:figs3a}(a), all mixtures are long-lived in the low-density MOT stage with lifetimes exceed 10~s except for $^{163}$Dy in the $^{163}$Dy--$^{164}$Dy mixture, which has a lifetime of only about 1~s, consistent with its strong sensitivity to the dual-frequency 626-nm configuration identified above.

To characterize the effect associated specifically with the presence of the second isotope, we define $R_{i,\tau}^{\rm MOT}=\tau_{i,\rm DDM}/\tau_{i,\rm DIS}$, which compares the lifetime of the full mixture with the corresponding single-isotope reference. As shown in Fig.~\ref{fig:figs3a}(b), $R_{i,\tau}^{\rm MOT}$ remains very close to unity for all isotope combinations despite the variations in their absolute lifetimes. This is also the case for $^{163}$Dy in the $^{163}$Dy--$^{164}$Dy mixture, where both DDM and DIS exhibit similarly short lifetimes. Thus, within our experimental sensitivity, introducing the second isotope produces no obvious additional loss holding in the low-density MOT. This result further supports that the pair-dependent loading reductions discussed above arise predominantly from optical effects during loading rather than from enhanced loss of the co-trapped mixture after capture.

After compression, however, the decay behavior becomes clearly non-exponential. We characterize the CMOT decay using a simple two-body-loss model, $N_i(t)=N_i(0)/[1+\kappa_iN_i(0)t]$, which captures the main features of the observed dynamics, with $N_i(0)$ and $\kappa_i$ treated as fit parameters. This model is used as an effective description and is not intended to imply purely two-body loss or to provide a quantitative microscopic two-body loss coefficient. Importantly, DIS provides the corresponding single-isotope reference, whereas DDM allows additional interspecies loss processes. The DDM--DIS comparison therefore enables us to distinguish additional interspecies-related loss from the single-isotope background, while $\kappa_i$ is used only as an effective parameter for this comparison.

The relative CMOT stability is therefore characterized by $R_{i,\kappa}^{\rm CMOT}=\kappa_{i,\rm DIS}/\kappa_{i,\rm DDM}$, with values below unity indicating faster effective decay in DDM. As the main distinction between the DDM and DIS cases is the presence of the second isotope, a reduced $R_{i,\kappa}^{\rm CMOT}$ provides an experimental signature of additional mixture-associated decay channels, including interspecies collisional loss. The ratio is nevertheless used as a comparative benchmark rather than a measurement of the interspecies loss coefficient. The systematic comparison across isotope combinations is presented in Fig.~\ref{fig:isotopes}(c) of the main text.

\begin{figure}[tb!]
	\includegraphics[width=0.9\columnwidth]{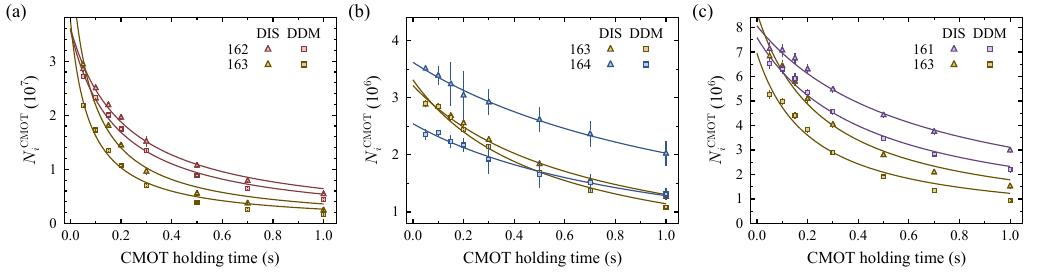}
	\caption{\textbf{CMOT decay dynamics for selected isotope combinations.}
		DDM and DIS decay dynamics of both components for (a) $^{162}$Dy--$^{163}$Dy, (b) $^{163}$Dy--$^{164}$Dy, and (c) $^{161}$Dy--$^{163}$Dy, corresponding to the combinations that exhibit pronounced deviations of $R_{i,\kappa}^{\rm CMOT}$ from unity in Fig.~5(c). Solid curves are fits to the simple two-body loss model. %$N_i(t)=N_i(0)/[1+\kappa_iN_i(0)t]$.
		}
	\label{fig:figs3b}
\end{figure}

Figure~\ref{fig:figs3b} shows the corresponding CMOT decay dynamics for the $^{162}$Dy--$^{163}$Dy, $^{163}$Dy--$^{164}$Dy, and $^{161}$Dy--$^{163}$Dy pairs, which exhibit the most pronounced deviations of $R_{i,\kappa}^{\rm CMOT}$ from unity in Fig.~\ref{fig:isotopes}(c). For all three combinations, both isotope components decay faster in DDM than in the corresponding DIS reference. The solid curves are fits using the same effective two-body-loss form, from which the corresponding $\kappa_i$ values are extracted for the DDM--DIS comparison. The enhanced decay of both components indicates additional mixture-specific collisional loss channels after the second isotope is introduced.

Taken together, the MOT- and CMOT-hold measurements reveal a clear change in mixture stability upon compression. In the dilute MOT, DDM and DIS exhibit closely matched lifetimes even when the absolute lifetimes differ among isotope combinations, indicating no substantial additional mixture-induced loss. At the higher CMOT density, however, $^{163}$Dy-containing mixtures exhibit obvious faster decay in DDM for both components. This behavior indicates additional interspecies collisional loss channels that exist only in the mixture. Because the fitted $\kappa_i$ remains an effective parameter that combines several contributions, the present measurements establish enhanced mixture-associated decay without assigning it to a specific microscopic interspecies loss coefficient.

\subsection{ODT transfer benchmarks}

To characterize ODT transfer across the different isotope combinations, we vary the vertical offset B field to shift the CMOT position relative to the ODT, with the alignments along the other two orthogonal directions fixed at their optimized values. For each isotope and preparation protocol, the resulting loading curve is fitted with a Gaussian function to determine the optimal loading position and peak atom number. We define the transfer efficiency as $\eta_i=N_{i}^{\rm ODT}/N_{i}^{\rm CMOT}$ and use the fitted loading maximum to obtain the peak efficiencies $\eta_{i,\rm DDM}^{\rm max}$ and $\eta_{i,\rm DIS}^{\rm max}$. The relative transfer performance is then characterized by $R_{i,\eta}^{\rm ODT}=\eta_{i,\rm DDM}^{\rm max}/\eta_{i,\rm DIS}^{\rm max}$, as summarized in Fig.~\ref{fig:isotopes}(d) of the main text. The CMOT and ODT atom numbers entering $\eta_i$ are obtained from independent measurements prepared under the corresponding experimental conditions rather than from shot-by-shot interleaved measurements.

\begin{figure}[tb!]
	\includegraphics[width=0.9\columnwidth]{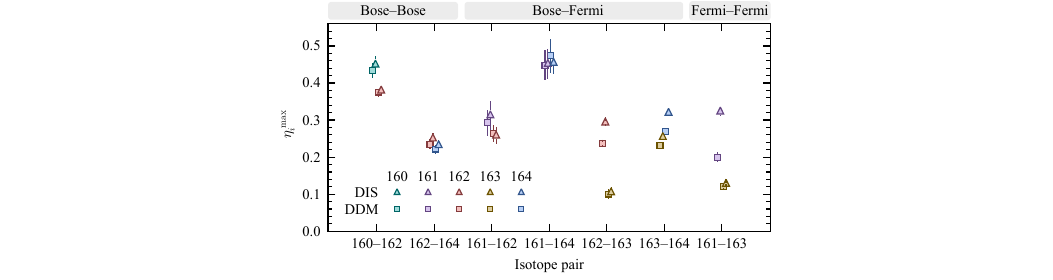}
	\caption{\textbf{Absolute ODT-transfer efficiencies across dysprosium isotope pairs.} Peak transfer efficiencies $\eta_{i,\rm DDM}^{\rm max}$ and $\eta_{i,\rm DIS}^{\rm max}$ for the different isotope combinations, obtained from Gaussian fits to the corresponding loading curves. The data provide the absolute efficiencies underlying the relative transfer ratios $R_{i,\eta}^{\rm ODT}$ shown in Fig.~\ref{fig:isotopes}(d). Colors denote isotope identity, while marker shapes distinguish DDM and DIS.}
	\label{fig:figs4}
\end{figure}

While $R_{i,\eta}^{\rm ODT}$ characterizes the relative change in transfer efficiency associated with dual-isotope operation, it does not specify the absolute transfer performance. We therefore show the corresponding absolute peak efficiencies $\eta_{i,\rm DDM}^{\rm max}$ and $\eta_{i,\rm DIS}^{\rm max}$ for all isotope combinations in Fig.~\ref{fig:figs4}. For each pair, the relative CMOT--ODT position is optimized through the offset B fields of all three directions. The other parameters involved in the CMOT-to-ODT loading, including the 626-nm laser parameters, magnetic-field gradient, and compression sequence, are not exhaustively re-optimized for each isotope combination. The absolute efficiencies therefore characterize the performance obtained with the common preparation sequence rather than the fully optimized transfer efficiencies attainable for individual pairs. Differences in the absolute values across isotope combinations should consequently not be interpreted as intrinsic limitations.

The complementary roles of the relative and absolute benchmarks are particularly clear for the $^{161}$Dy--$^{163}$Dy combination. The $^{163}$Dy component exhibits a relatively low peak transfer efficiency in both DDM and DIS, while $R_{163,\eta}^{\rm ODT}$ remains close to unity. Its low absolute efficiency is therefore already present in the corresponding DIS reference under the preparation conditions used here rather than arising primarily from dual-isotope operation. In contrast, the $^{161}$Dy component exhibits a DIS transfer efficiency comparable to those of most other components but a substantially lower value in DDM, resulting in the pronounced reduction of $R_{161,\eta}^{\rm ODT}$ in Fig.~\ref{fig:isotopes}(d).

Taken together, the relative ratios in Fig.~\ref{fig:isotopes}(d) and the absolute peak efficiencies in Fig.~\ref{fig:figs4} provide complementary measures of the ODT-transfer performance. The former quantify the additional change associated with dual-isotope operation under otherwise matched loading conditions, whereas the latter establish the single-isotope performance reached with the common preparation sequence. For most components, the DDM peak efficiencies remain close to their DIS references, supporting nearly single-isotope-like ODT-transfer performance across the explored isotope combinations, while the lower absolute efficiencies of selected components leave room for further pair-specific optimization.

\end{document}